\documentclass[aps,prl,superscriptaddress,twocolumn]{revtex4-1}
\usepackage{graphicx}
\usepackage{amsmath,amssymb,amsfonts}
\usepackage{hyperref}
\usepackage{color}
\usepackage{natbib}
\usepackage{changes}
\usepackage[english]{babel}
\def \EPFLString {Institute  of  Physics, \'{E}cole Polytechnique F\'{e}d\'{e}rale de Lausanne (EPFL), Lausanne, CH-1015, Switzerland}
\def \UCAMString {Cambridge Graphene Centre, University of Cambridge, Cambridge CB3 0FA, UK}
\def \JHUString {Department of Electrical and Computer Engineering, Johns Hopkins University, Baltimore, Maryland 21218, USA}
\def \SharifString {School of Electrical Engineering, Sharif University of Technology, Tehran, Iran}
\begin{document}
\title{Excitonic emission of monolayer semiconductors near-field coupled to high-Q microresonators}
\author{Cl\'{e}ment Javerzac-Galy}
\thanks{These authors contributed equally to this work}
\affiliation{\EPFLString}
\author{Anshuman Kumar}
\thanks{These authors contributed equally to this work}
\affiliation{\EPFLString}
\author{Ryan D. Schilling}
\affiliation{\EPFLString}
\author{Nicolas Piro}
\affiliation{\EPFLString}
\author{Sina Khorasani}
\affiliation{\SharifString}
\affiliation{\EPFLString}
\author{Matteo Barbone}
\affiliation{\UCAMString}
\author{Ilya Goykhman}
\affiliation{\UCAMString}
\author{Jacob B. Khurgin}
\affiliation{\JHUString}
\author{Andrea C. Ferrari}
\affiliation{\UCAMString}
\author{Tobias J. Kippenberg}
\affiliation{\EPFLString}
\email{tobias.kippenberg@epfl.ch}
\begin{abstract}
We present quantum yield measurements of single layer $\textrm{WSe}_2$  (1L-$\textrm{WSe}_2$) integrated with high-Q ($Q>10^6$) optical microdisk cavities, using an efficient ($\eta>$90\%) near-field coupling scheme based on a tapered optical fiber. Coupling of the excitonic emission is achieved by placing 1L-WSe$_2$ to the evanescent cavity field. This preserves the microresonator high intrinsic quality factor ($Q>10^6$) below the bandgap of 1L-WSe$_2$. The nonlinear excitation power dependence of the cavity quantum yield is in agreement with an exciton-exciton annihilation model. The cavity quantum yield is $\textrm{QY}_\textrm{c}\sim10^{-3}$, consistent with operation in the \textit{broad emitter} regime (i.e. the emission lifetime of 1L-WSe$_2$ is significantly shorter than the bare cavity decay time). This scheme can serve as a precise measurement tool for the excitonic emission of layered materials into cavity modes, for both in plane and out of plane excitation.
\end{abstract}
\maketitle
\begin{figure*}
\centerline{\includegraphics[width=180mm]{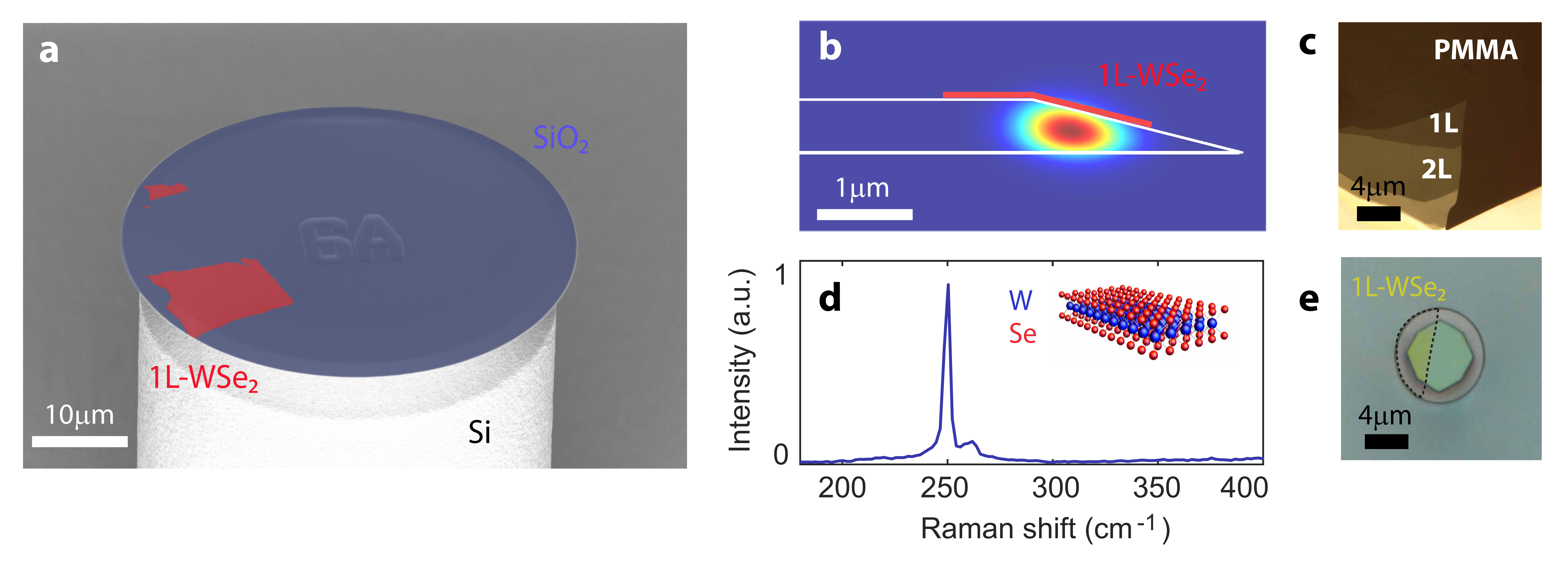}}
\caption{(a)False-colored scanning electron micrograph of a 750~nm thick SiO$_2$ (blue) microdisk integrated with 1L-WSe$_2$ (red). The microdisk has a radius of 19~$\mu$m and sits on a $\sim200$~$\mu$m Si pillar to prevent the tapered fiber from being in contact with the substrate  (b) Finite element simulation showing that the WGM transverse mode profile is located within the wedge. 1L-WSe$_2$ lies in the near-field of the optical mode. (c) Optical image showing the contrast of the various layers of exfoliated WSe$_2$ prior to deposition on the microdisk. (d) Raman spectrum of the microdisk integrated 1L-WSe$_2$. (e) Optical image of a microdisk integrated 1L-WSe$_2$ (radius $\sim8$~$\mu m$) showing almost complete coverage (dotted line).}
\label{fig:SEM}
\end{figure*}
Transition metal dichalcogenides (TMDs) are promising for opto-electronic applications\cite{Nanoscale2015}, including light emitting devices\cite{10.1038/nmat4205}, optical modulators\cite{/10.1038/nphoton.2016.15} and photo-detectors\cite{Koppens2014}. These exploit the fact that their optical response is dominated by strong excitonic transitions, with large binding energies of a few hundred meV\cite{PhysRevB.92.245123,PhysRevB.86.115409} as a result of the reduced screening. The broken inversion symmetry, in combination with the strong spin-orbit coupling leads to the possibility of valley optoelectronic devices\cite{10.1038/natrevmats.2016.55}. Exfoliated TMDs have a direct bandgap\cite{PhysRevLett.105.136805}. They could be integrated in cavities with the aim of enhancing the light matter interaction via the increased optical density of states\cite{10.1038/nphoton.2014.271,Mak2016}. Enhanced interactions have been demonstrated with layered materials (LMs) integrated with photonic crystal cavities\cite{doi:10.1063/1.4826679,2053-1583-1-1-011001}, distributed Bragg cavities\cite{10.1038/nphoton.2014.304,Dufferwiel2015} and microdisks\cite{10.1038/nphoton.2015.197}. A key metric of light emission in such integrated systems is the \emph{cavity quantum yield} (QY$_c$), i.e. the number of photons emitted into the cavity mode for each absorbed pump photon. The QY$_c$ is often limited by the non-radiative recombination of excitons in TMDs, which mostly arises from defects and Auger effects\cite{Amani1065,C5NR00383K,doi:10.1021/nl503636c}. An accurate QY$_c$ measurement tool is needed to attain an improved physical understanding of emission properties in these systems, and for evaluation of future opto-electronic devices performance, such as their efficiency or expected output power. Conventional techniques for QY$_c$ measurement rely either on free space excitation and collection\cite{10.1038/nprot.2013.087} or fitting the photoluminescence (PL) decay rate as a function of a geometrical parameter of the cavity\cite{chizhik2013nanocavity}. These techniques are usually not suitable for cavity integrated device measurements as they do not collect the light directly from the cavity mode.

A precise QY$_c$ characterization is also needed to establish lasing operation in cavity integrated LMs, that has recently captured significant interest\cite{Mak2016,10.1038/nature14290,10.1038/nphoton.2015.197,Li2017}. In general, the determination of the lasing threshold in a microlaser is compounded\cite{Samuel2009} by the fact that microlasers can exhibit a large (close to unity) spontaneous emission coupling factor ($\beta$), defined as the fraction of total spontaneous emission that is captured into the cavity mode, as typical for devices with a small wavelength-size mode volume. For emitters spectrally broader than the cavity mode (i.e. which decay significantly faster than the bare cavity), the Purcell effect, i.e. the enhancement of spontaneous emission rate by a cavity\cite{PhysRev.69.674,Haroche1989}, is dominated by the emitter linewidth. Specifically in this \textit{broad emitter} regime, also referred to as the \textit{bad emitter} regime in cavity quantum electrodynamics (cQED), a large (close to 1) $\beta$ cannot be achieved, and the achievable enhancement is governed by both emitter linewidth and cavity linewidth\cite{PhysRevB.81.245419}. This regime also applies to Refs.\citenum{10.1038/nature14290,10.1038/nphoton.2015.197,Li2017}, which consider LM coupled to nanophotonic cavities, where the emission was interpreted as lasing. However, the inferred $\beta$ value in these studies (i.e. $\beta\sim$ 0.1 to $\sim$ 0.5) are not expected, and in fact incompatible, with the broad emitter regime.

Importantly, regardless of the Purcell effect, the transition from spontaneous to stimulated emission dominated regime occurs in \emph{any laser}, irrespective of implementation, when the mean photon number in the cavity lasing mode exceeds unity ($\bar{n}_\text{c}>1$). The underlying reason is that stimulated emission and spontaneous emission into the cavity differ only by the bosonic mode occupation of the lasing mode\cite{MilonniBook}. This condition has also been referred to as the \emph{quantum threshold condition}\cite{PhysRevA.50.1675} in the context of early studies of microlasers.

This requirement implies (see Methods) that the lowest possible threshold to achieve a unity cavity photon number in the lasing mode (for an optically pumped ideal single mode laser) $G_{\text{th,ideal}}$ is given by a pump flux that equals the cavity decay rate\cite{PhysRevA.50.1675}. In fact for an optically pumped microlaser, assuming $\beta=1$ and neglecting non-radiative losses, the pumping rate of the cavity mode given by $G_{\text{th}}=P_{\textrm{th}}/\hbar\omega_{\textrm{p}}$, with $P_{\textrm{th}}$ the threshold pump power, $\hbar\omega_{\textrm{p}}$ the pump photon energy, should equal or exceed the cavity decay rate given by $\kappa_\text{tot} = \omega_{\textrm{lasing}}/Q$, with $\omega_{\textrm{lasing}}$ the lasing angular frequency and $Q$ the loaded quality factor at the lasing frequency. For $\beta=1$ and an ideal (lossless) emitter, the lowest possible threshold depends only on the total cavity loss rate $\kappa_\text{tot}$.

This general condition is satisfied in a variety of lasers at the micro- and nano-scale, such as photonic crystal defect lasers coupled to quantum wells\cite{Painter1819} ($G_{\text{th}}\approx 3\times 10^{16}~\text{s}^{-1}>G_{\text{th,ideal}}\approx\kappa_\text{tot}\approx 6\times 10^{12}~\text{s}^{-1}$, where non-radiative channels are negligible\cite{Painter1819}) and rare earth-doped microdisk lasers\cite{PhysRevA.74.051802} ($G_{\text{th}}\approx3\times10^{14}~\text{s}^{-1}>G_{\text{th,ideal}}\approx6\times10^9~\text{s}^{-1}$). However, in Ref.\citenum{10.1038/nature14290}, where lasing from 1L-WSe$_2$ was claimed, this condition is \textit{not fulfilled}: $G_{\text{th}}\approx1\times10^{11}~\text{s}^{-1}<G_{\text{th,ideal}}\approx1\times10^{12}~\text{s}^{-1}$, even \textit{neglecting} non-radiative relaxation processes. More generally there are growing concerns on rigorous identification of lasing\cite{10.1038/nphoton.2017.28}. The increasing research effort on LMs and heterostructures of interest for nano-scale light sources needs a precise technique which can characterize the exact number of photons absorbed and emitted, to verify if the threshold condition for lasing is reachable or met.

Here, we measure and characterize the optical response of cavity integrated TMDs (and the exciton emission) by using a tapered fiber coupling scheme\cite{PhysRevA.74.051802} to both pump 1L-TMD and collect the emission, thereby determining QY$_c$ upon optical pumping. This method enables the determination of the exact power absorbed by 1L-TMD, as well as the photons emitted from it into the cavity modes, via the determination of the external (taper fiber induced) and intrinsic cavity decay rates. This allows us to calculate precisely the QY$_c$. Moreover, tapered fiber excitation of the TMD integrated microresonator allows us to excite both in plane and out of plane polarization modes, in principle enabling selective excitation of dark and bright excitons, which couple to distinct polarizations\cite{2017arXiv170405341W,2017arXiv170105938Z}. One may let one of the two available orthogonal polarizations at each wavelength enter the resonator and interact with the LM. We demonstrate our approach by measuring QY$_c$ of 1L-WSe$_2$ integrated with a silica microdisk resonator. We show that this maintains high intrinsic Q ($>10^6$ below gap, corresponding to an optical Finesse exceeding $10^4$), higher than in previous reports\cite{10.1038/nature14290,10.1038/nphoton.2015.197}, enabling the fabrication of optoelectronic devices without need of heterogeneous integration\cite{Heterogeneousintegration}.

The system under study consists of a SiO$_2$ microdisk, with a 1L-$\textrm{WSe}_2$ flake deposited on its surface. This is prepared as follows. A 750~nm thick $\textrm{SiO}_2$ film is grown on a Si(100) wafer by thermal oxidation. Wet etching in buffered hydrofluoric acid forms the microdisks\cite{PhysRevApplied.5.054019}. The wet etch mask is defined by electron beam lithography in ZEP 520A polymeric photo-resist. A post development bake is performed to reduce stress in the polymer and increase adhesion. The resulting microdisks have diameters from $\sim$16 to $\sim$38~$\mu$m. They have a sidewall wedge angle of 10$^\circ$, characteristic of such wet etching processes\cite{GHULINYAN2015679}, which serves two purposes: first, as shown in Fig.\ref{fig:SEM}, it results in the mode being shifted away from the disk edge, thereby reducing scattering losses\cite{PhysRevA.74.051802}. Secondly, it enables tapered fiber coupling in contact mode (away from the mode center), thereby reducing cavity power fluctuations due to taper vibrations. Finally, the microdisks are undercut with potassium hydroxide to form an air-clad whispering gallery mode (WGM) resonator\cite{10.1038/nature01939}. The microdisks sit on $\sim200$~$\mu$m Si Mesa pillars to prevent the tapered fiber from being in contact with the substrate during coupling. The microdisks undergo an oxygen plasma cleaning before 1L-$\textrm{WSe}_2$ transfer.

\begin{figure*}
\includegraphics[width=160mm]{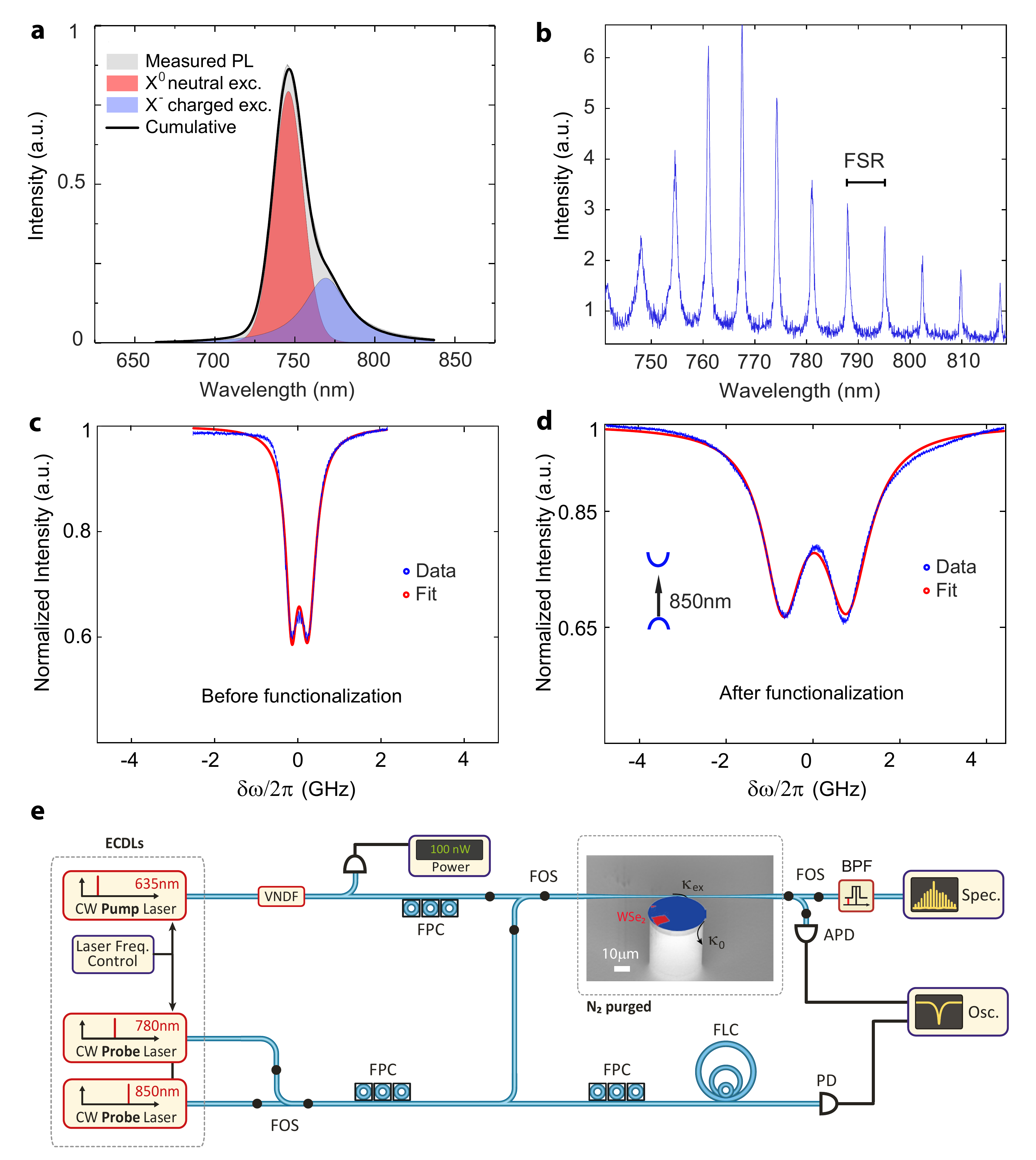}
\caption{(a) Room-temperature PL of microdisk integrated 1L-WSe$_2$ shows neutral and charged exciton contributions. (b) Cavity enhanced PL of microdisk integrated 1L-WSe$_2$ with background-free emission channeled into the WGMs. Excitation and collection are both performed via a tapered fiber. The FSR for this microdisk is $\sim$ 8~nm. (c,d) Linewidths of bare and 1L-WSe$_2$ integrated microdisks at 850~nm (below gap, as depicted in the inset). The measured loaded quality factors are 9$\times 10^5$ (bare) and 3$\times 10^5$ (1L-WSe$_2$ integrated). Nanoscale surface defects such as a small imperfections cause the high-Q resonances to split. (e) Schematic setup for precise characterization of emission efficiency. This uses a two color scheme where the pump laser is coupled into the microdisk via the near-field of a tapered fiber with coupling rate $\kappa_{\textrm{ex}}$ and the emitted light is collected by the same tapered fiber. Probe lasers are used to characterize the 1L-WSe$_2$ integration and the coupling of the microdisk at different wavelengths around the 1L-WSe$_2$ bandgap. The intrinsic loss rate is represented by $\kappa_{0}$. ECDLs: External Cavity Diode Lasers, VNDF: Variable Neutral Density Filter, FOS: MEMS Fiber Optical Switch, FPC: Fiber Polarization Controller, FLC: Fiber Loop Cavity, APD: Avalanche Photo-diode, PD: Photo-diode, BPF: Bandpass Filter, Spec.: Optical grating spectrometer, Osc.: Oscilloscope.}
\label{fig:setup}
\end{figure*}

1L-WSe$_2$ is prepared and integrated on the microdisks as follows. Bulk WSe$_2$ (HQ Graphene) is characterized before exfoliation by Raman Spectroscopy as described in\cite{10.1038/ncomms12978}. This is then exfoliated on a polydimethylsiloxane layer by micromechanical cleavage. 1L-WSe$_2$ flakes samples are identified by optical contrast\cite{10.1021/nl071168m}. Selected 1L-WSe$_2$ flakes are then transferred onto the microdisks via an all-dry viscoelastic transfer technique, exploiting their higher adhesion to SiO$_2$\cite{2053-1583-1-1-011002}. After transfer, the flakes are characterized by Raman spectroscopy (Fig.\ref{fig:SEM}(d)) at 532~nm excitation. The main features are the A1'+E' mode at $\sim$ 249.5~cm$^{-1}$ and the 2LA(M)\cite{doi:10.1038/srep04215}. The thickness is then confirmed by PL (Fig.\ref{fig:setup}(a))\cite{10.1039/C3NR03052K}. This confirms the transfer and that the process does not damage the samples.

A scanning electron microscopy (SEM) image of the 1L-WSe$_2$ integrated microdisk is shown in Fig.\ref{fig:SEM}(a). The WGM transverse mode profile is located within the wedge, as indicated by the finite element simulation (using Comsol Multiphysics) in Fig.\ref{fig:SEM}(b). Thus, the 1L-WSe$_2$ flake sits in the near-field of the optical mode. The 1L-WSe$_2$ is modeled as a $\sim0.65$~nm thick dielectric\cite{10.1103/PhysRevB.90.205422}.

In order to allow excitation and collection of emission from the 1L-WSe$_2$, we use a near field coupling scheme via a tapered optical fiber\cite{PhysRevLett.91.043902}, whereby phase matching between the WGM and the tapered fiber mode is obtained by translating the disk along the taper waist\cite{PhysRevLett.85.74}. For a coupling ideality of unity\cite{PhysRevLett.91.043902}, the coupling parameter is given by\cite{Haus1984}:
 \begin{equation}
\frac{\kappa_{\mathrm{ex}}}{\kappa_{0}}=\frac{1 \pm \sqrt{T}}{1 \mp \sqrt{T}},
\label{eq:kappa}
 \end{equation}
\noindent
where $\kappa_{0}$ is the intrinsic loss rate of the microresonator, which is the sum of radiative and absorption losses. The latter are dominated by 1L-WSe$_2$ for wavelengths above the bandgap because of interband absorption. Here, $\kappa_{\mathrm{ex}}$ is the photon loss rate due to the external coupling of light to the tapered fiber, and $T$ the transmission of the microresonator on resonance. The upper signs are used for the over-coupled regime ($\kappa_{0}<\kappa_{\mathrm{ex}}$) and the lower signs for the under-coupled regime ($\kappa_{0}>\kappa_{\mathrm{ex}}$)\cite{Haus1984}. T vanishes for critical coupling ($\kappa_0=\kappa_{\mathrm{ex}}$)\cite{PhysRevLett.85.74}.

We then record the transmission spectrum while scanning an external cavity diode laser (ECDL) over resonance for different central wavelengths and taper waist radii. An ECDL is necessary because we need mode-hop free excitation over the free spectral range, which is typically sub THz. The corresponding total linewidth ($\kappa_{\textrm{tot}} = \kappa_{\textrm{ext}}+\kappa_0$) is measured with a calibrated fiber loop cavity. Thus with the knowledge of $\kappa_{\textrm{tot}}$ and $T$, the external and intrinsic coupling rates, $\kappa_{\textrm{ext}}$ and $\kappa_0$, are determined. This is one of the key features of our approach, which enables precise characterization of photon emission and absorption rates of LMs.

Our characterization setup is presented schematically in Fig.\ref{fig:setup}. Laser light through the ECDL is split into two parts: one passes through a fiber loop cavity (calibration branch) and another through the tapered fiber, which is coupled to the cavity (cavity branch). Both branches have independent fiber polarization controllers. For linewidth calibration, the reference and the cavity branch are detected and are monitored on an oscilloscope. The free spectral range of the fiber loop cavity is known through an independent calibration via a phase modulation measurement. In our setup, the fiber loop cavity serves to calibrate the time axis on the oscilloscope. In this way, the exact total linewidth and the transmission resonance is recorded automatically. For measurement of the spectrum, the output of the cavity branch is sent to a spectrometer. A high pass filter (cut-on wavelength of 660~nm) is employed prior to the spectrometer to cancel out the pump.

We first characterize the bare microdisks in a one-color scheme, where we excite and collect at the same laser excitation wavelength. We use 3 excitation wavelengths: below gap (850~nm), near excitonic transition (770~nm) and above gap (635~nm). We consistently observed Q factors $>5\times 10^5$ for the \emph{bare} disks at all wavelengths. An example transmission spectrum for the bare disk is plotted in Fig.\ref{fig:setup}(c) for $\lambda=850$~nm, where a double Lorentzian fit leads to a $Q_0=\omega/\kappa_0\sim1\times 10^7$. Nano-scale surface defects such as a small imperfections cause the high-Q resonances to split\cite{Weiss:95} due to the coupling between the clockwise and counterclockwise propagating WGMs.

Next, we measure the linewidths for the 1L-WSe$_2$ coated disks. As shown in Fig.~\ref{fig:setup}(d), the intrinsic $Q_0$ is maintained at a high value (e.g. $4\times 10^6$ at 850~nm). We attribute the small reduction in $Q_0$ at this wavelength to scattering at the 1L-WSe$_2$ edges.

We note that our disk platform provides Q ($>10^6$) and Finesse ($>10^4$) below gap, exceeding previous reports of cavity integrated LMs\cite{doi:10.1021/acs.nanolett.5b01665,10.1038/nature14290,10.1038/nphoton.2015.197,doi:10.1021/nl503312x,10.1038/nphoton.2014.304}. For excitations above the bandgap, the Q ($\sim 10^3$) and Finesse ($\sim 10$) are reduced, demonstrating absorption dominated behavior due to the 1L-WSe$_2$.

The near field coupling to the microdisk via a tapered fiber enables in principle a precise determination of the photons absorbed as well as those emitted into the cavity modes, when the loss rate at the pump wavelength is dominated by the LM absorption. Our method can be employed to characterize $\textrm{QY}_c=R_{\textrm{cav,exc}} /G$, where $R_{\textrm{cav,exc}}$ is the emission rate into the cavity modes and $G$ is the pumping rate.

The 1L-WSe$_2$ is excited using a tunable pump laser at $\lambda=635~\textrm{nm}$ via near field coupling using a tapered fiber. The pump frequency is tuned to a resonance of the WGM of the microdisk. The resulting 1L-WSe$_2$ PL (at~$\sim760$~nm) is coupled into the WGM of the microdisk near the 1L-WSe$_2$ A-exciton transition energy\cite{PhysRevLett.113.026803,PhysRevLett.105.136805}. The exciton emission coupled to the WGM is then collected again via the same near field coupled tapered fiber. Thus, unlike the pristine microdisk case, this measurement employs a two color scheme, where the pump and emission wavelengths are different.

Fig.\ref{fig:setup}(b) indicates that the resulting emission consists of background-free cavity-enhanced PL with several peaks arising from the WGMs of the microdisk. Refs.\citenum{doi:10.1021/acs.nanolett.5b01665,10.1038/nphoton.2015.197} reported emission measurement schemes on cavity integrated LMs, where the cavity modes are superimposed on a background of the free-space PL\cite{10.1038/nphoton.2015.197}. Our near field tapered fiber scheme enables collection only of the PL which is coupled into the cavity modes. Thus, without background PL arising from coupling into free space modes. The PL collection efficiency in our setup is defined as $\eta=\kappa_{\textrm{ext}}/\kappa_{\textrm{tot}}$, which represents the collection efficiency of the photons emitted into the cavity modes\cite{Grillet:07}. Unlike free space collection techniques, we are not limited by the numerical aperture of the objective and under appropriate phase matching, in principle $\eta>90\%$ is achievable, when operating strongly overcoupled\cite{PhysRevLett.85.74}.

The peaks in our taper collected PL are spaced by the free spectral range (FSR) that characterizes the separation of two longitudinal resonances of the disk: $FSR=\lambda^2/(2\pi n_{\textrm{g}} R)$, where $\lambda$ is the central wavelength, $n_{\textrm{g}}$ is the effective mode index and $R$ is the microdisk radius. In Fig.\ref{fig:setup}(b) the FSR is $\sim8$~nm, as expected from $R\sim 8~\mu m$. Secondary peaks might arise from either fundamental modes of the orthogonal polarization or from higher order modes of the microdisk.

We observe an asymmetric lineshape in the free space PL which is well fitted to the sum of two Voigt peaks, arising from the neutral and charged excitons\cite{PSSB:PSSB201600563}. Although the rigorous identification of the lower energy peak would require a doping-dependent PL measurement, it is usually attributed to trionic emission\cite{PhysRevB.90.075413,doi:10.1063/1.4895471}. Thus we attribute the asymmetric shape of the envelope of our cavity PL to the spontaneous emission profile of the 1L-WSe$_2$. 

\begin{figure*}
\includegraphics[width=\textwidth]{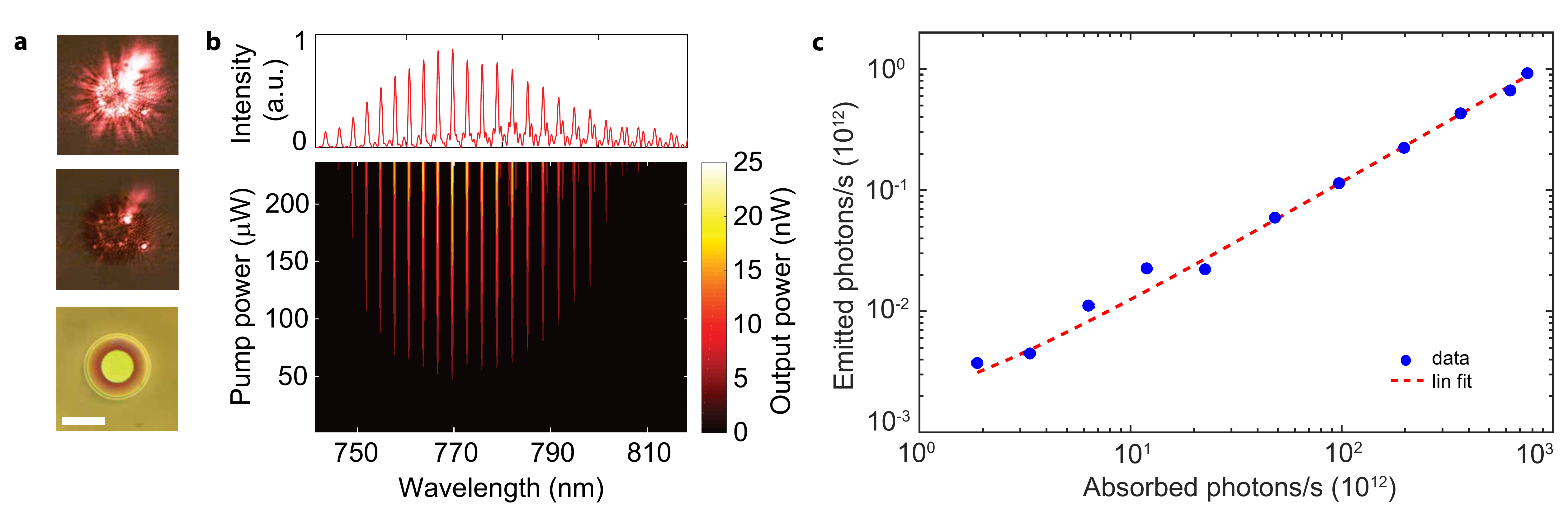}
\caption{(a) Scattered light from the 1L-WSe$_2$ on the microdisk as the excitation power is increased from $\sim0.5~\mu W$ to $\sim250~\mu W$ (bottom to top). Scalebar 20~$\mu$m. (b) Excitation power dependence of the cavity enhanced PL collected via a tapered fiber at room temperature. (c) QY$_c$ measurement using the tapered fiber technique. A linear fit to the photons absorbed and emitted into all longitudinal cavity modes yields QY$_c\sim$ 0.1\% after correcting for the coupling efficiency at the pump and emission wavelengths.}
\label{fig:PL}
\end{figure*}

Fig.\ref{fig:PL}(a) shows the enhancement of PL from 1L-WSe$_2$ integrated microdisk as the excitation power is increased from $\sim0.5~\mu W$ to $\sim250~\mu W$. Scattered light of the PL coupled into the WGM can be observed. In Fig.\ref{fig:PL}(b), $FSR\simeq$ 3~nm and $\eta\approx 93\%$. This scattered PL is a fraction (see Methods) of the total emission, the rest going into the WGMs and non-radiative channels. Fig.\ref{fig:PL}(b) shows the shape of the PL spectrum as a function of the excitation power.

Our approach for measuring linewidths and $T$ allows us to determine the values of the internal and external coupling rates at both pump and emission wavelengths. This enables us to calibrate the pump power input as well as the emission output. Thus, this technique enables precise determination of the QY$_c$ of cavity-integrated LMs. The data in Fig.\ref{fig:PL}(c) includes this calibration, giving QY$_c\sim0.10\%$ at room temperature.

When an emitter is placed near a cavity, its emission rate into the cavity mode is modified compared to free space. When this rate is enhanced, it is referred to as Purcell enhancement\cite{purcell1995spontaneous}. For a high-Q cavity this can give a QY$_c$ increase\cite{PhysRevLett.116.247402}. The simulated \emph{peak} Purcell enhancement factor for each cavity mode is\cite{10.1038/nature01939}:
\begin{equation}
F_p = (3/4\pi^2) (\lambda_0 / n)^3 Q_c/V_{\textrm{eff}}
\label{purcell}
\end{equation}
where $\lambda_0 / n$ is the wavelength within the material and, $Q_c$ and $V_{\textrm{eff}}$ are the quality factor and mode volume of the cavity, respectively. This is $\sim$ 8-14 for the different mode families within our window of operation. However, Eq.\ref{purcell} is only applicable when the cavity linewidth dominates that of the emitter\cite{refId0}. This explains why no Purcell enhancement is observed in our experiment, since our cavity linewidth is $<0.01$~nm and the emitter linewidth is $>10$~nm. The total emission rate is determined by the integral of the product of the spectral density of the excited states in the emitter and the photon density of states. This can be understood in terms of an approximate expression for the Purcell factor (on resonance), which for one of the cavity modes $i$ is given by\cite{Meldrum:10,refId0}:
\begin{equation}
F_{p,i} = \frac{3}{4\pi^2}\frac{(\lambda/n)^3}{V_{\textrm{eff}}} \left(\frac{1}{ Q_{\textrm{em}} } + \frac{1}{ Q_{c,i} }\right)^{-1}
\end{equation}
where, ${Q_{c,i}}$ and $Q_{\textrm{em}}$ are the quality factors of the cavity mode $i$ and the emitter. Due to the large emission linewidth $(\sim 50~\textrm{meV})$ (see Fig.\ref{fig:setup}) of 1L-WSe$_2$ compared to the cavity modes, the system is in the \textit{broad emitter} regime. This leads to a reduction in the Purcell factor limiting the spontaneous emission enhancement.

The error in the linearly fitted QY$_c$ (without taking into account the error in the data collected), as presented in Fig.\ref{fig:PL}(c), is $\sim$ 5\% . This is attributed to the fact that the QY$_c$ of the 1L-WSe$_2$  depends on pump intensity, similar to what was observed for exfoliated 1L-MoS$_2$\cite{Amani1065}. Thus, any linear fit with varying pump power is expected to produce errors. We present intensity dependent QY$_c$ measurements in the Methods section. The degradation of the QY$_c$ with increasing pump power is attributed to exciton-exciton annihilation (EEA)\cite{PhysRevB.90.155449}. We fit (See Methods) the QY$_c$ power dependence to a model taking both EEA as well as defect mediated non-radiative recombination into account. We consider a generation-recombination rate balance:
\begin{flalign}
G  &= R_{\text{rad,exc}} + R_{\text{trap,exc}} + R_{\text{EEA,exc}} \nonumber \\
& =  \tau_{\text{r}}^{-1}\langle N\rangle +\tau_{\text{nr}}^{-1}\langle N\rangle + \gamma_{\text{bx}}\langle N\rangle^2  , \nonumber
\end{flalign}
where $\langle N\rangle$ is the number of excitons, $\tau_{\text{nr}}$ is the non-radiative recombination time and $\gamma_{\text{bx}}$ is the EEA rate. The radiative recombination rate $\tau_{\text{r}}^{-1}\langle N\rangle$ is the sum of emission into two channels: free space and cavity modes. Thus, $\tau_{\text{r}}^{-1} = \tau_{\text{fs}}^{-1} + \sum\tau_{\text{c}}^{-1} = \tau_{\text{fs}}^{-1} + \tau_{\text{c,tot}}^{-1}$. We then get\cite{doi:10.1063/1.4978868}:
\begin{equation}
\textrm{QY}_c = \frac{R_{\textrm{cav,exc}} }{G} =\frac{\tau_{\text{c,tot}}^{-1}\langle N\rangle}{\tau_{\text{r}}^{-1}\langle N\rangle +\tau_{\text{nr}}^{-1}\langle N\rangle  + \gamma_{\text{bx}}\langle N\rangle^2}
\label{eq:QYc}
\end{equation}
We can rewrite Eq.\ref{eq:QYc} in a more physically intuitive form as:
\begin{equation}
\textrm{QY}_c(P) = \eta_{\textrm{c}}(0)\cdot\eta_{\textrm{EEA}}(P/P_0)
\end{equation}
\noindent
where $\eta_{\textrm{c}}(0)$ is QY$_c$ at low exciton densities and $\eta_{\textrm{EEA}}(P/P_0)$ is the QY$_c$ reduction caused by EEA, with $P_0$ being the associated power scale (See Methods). At high power ($>65~\mu W$ in our case), the model predicts a nonlinear dependence of QY$_c$ on pump power. This will limit the efficiency of light sources based on 1L-WSe$_2$.

We next ascertain the feasibility of achieving lasing. The low $QY_c\sim0.10$\% is insufficient for a given pump power to reach a single intra-cavity photon per mode ($\bar{n}_\text{c} \approx G \cdot  QY_\text{c}/(M \cdot\kappa_\text{tot})\approx 0.6$ at the highest pump power$\sim 250~\mu W$), as required for the quantum threshold lasing condition discussed earlier. This explains why no narrowing of the emission lines is observed. To achieve lasing, QY$_c$ must be increased by at least one order of magnitude by reducing non-radiative recombinations\cite{Amani1065}, as well as increasing $Q_{\textrm{em}}$, which contributes to the cavity decay rate, QY$_c$  and the number of cavity modes $M$ (see Methods). Factors such as $\beta$ and Purcell enhancement do not help reducing the threshold. While increasing $\beta$ is often presented as a means to reduce the laser threshold, this is almost always achieved by reducing the active volume of the laser\cite{Khurgin2014}, which, however keeps the all-important pumping density per unit of volume constant. Similarly, the Purcell enhancement does not assist as both stimulated and spontaneous emission rate increase by the same amount\cite{Khurgin2014,MilonniBook}

By optimizing the mode volume, our platform can be designed to show high ($>1$) effective Purcell factors for enhancing the QY$_c$ of LMs\cite{PhysRevLett.116.247402}. In addition, methods have been reported in literature to increase the LM QY, e.g. via encapsulation\cite{10.1088/2053-1583/aa6aa1} or chemical surface treatment\cite{Amani1065}.

In summary, we presented a technique for measuring the cavity coupled excitonic emission of high-Q microresonators integrated with layered materials, using a near-field coupled tapered fiber and a LM-integrated high-Q microdisk. We presented QY$_c$ measurements of 1L-WSe$_2$, obtaining $\sim10^{-3}$, in agreement with operation in the \textit{broad emitter} regime, i.e. the emission lifetime of 1L-WSe$_2$ significantly exceeds the cavity decay rate. These results contrast previous work on LM coupled to microcavities\cite{10.1038/nature14290,10.1038/nphoton.2015.197,Li2017}, that did not consider this regime, which applies equally to nanophotonic cavities due to the short excitonic emission lifetime of the LM. The low cavity quantum yields in the present case prevents reaching the threshold condition of lasing. Moreover, we studied the QY$_c$ excitation power dependence, and showed this it is well explained by a model based on exciton-exciton annihilation that limits the efficiency of light sources based on 1L-WSe$_2$\cite{10.1103/PhysRevB.93.201111}. Our approach can provide a route to standardization for LMs and their heterostructures, to compare their emission efficiency for device applications. For studying the fundamental physics of excitons in LMs, our technique is attractive since it allows for precise polarization control and near field mode engineering both for excitation as well as collection, which is useful for studying phenomena such as dark excitons\cite{2017arXiv170405341W,2017arXiv170105938Z} as well as enhanced light matter interaction\cite{10.1038/nphoton.2014.304,10.1038/nature01939}.
\section*{Acknowledgements}
We thank Giancarlo Soavi, Gordon Callsen, Mitchell Anderson, Eli Kapon, Itay Shomroni and Liu Qiu for useful discussions. We acknowledge funding from the EU Graphene Flagship, ERC Grant Hetero2D, EPSRC Grants EP/L016087/1, EP/K017144/1, EP/K01711X/1, the Swiss State Secretariat for Education, Research and Innovation (SERI ESKAS-Nr. 2016.0430), the Research Deputy of Sharif University of Technology. Samples were fabricated in the Center of MicroNanoTechnology (CMi) at EPFL.
\section*{Methods}
\subsection{Quantum threshold criterion}
For a single-mode laser, the rate equations for the mean exciton number $\langle N \rangle$ and the mean cavity photon number in a particular mode $\langle n_\text{c} \rangle$ in our semiconductor monolayer can be written as\cite{PhysRevA.50.1675}:
\begin{align}
\dfrac{\text{d}\langle N \rangle}{\text{d}t} &= G - \dfrac{\langle N \rangle}{\tau_{\text{r}}} -\dfrac{\beta \langle n_\text{c} \rangle}{\tau_{\text{r}}}\langle N \rangle  \\
&- \dfrac{\langle N \rangle}{\tau_{\text{nr}}} - \gamma_{bx}\langle N \rangle^2 \nonumber\\
\dfrac{\text{d}\langle n_\text{c} \rangle}{\text{d}t} &= - \kappa_{\text{tot}} \langle n_\text{c} \rangle + \dfrac{\beta}{\tau_{\text{r}}}\langle N \rangle + \dfrac{\beta \langle n_\text{c} \rangle}{\tau_{\text{r}}}\langle N \rangle
\label{eq:rateEq}
\end{align}
where $G$ is the pumping rate, $\beta$ is the spontaneous emission factor into the cavity mode ($\beta = \tau_\textrm{c}^{-1}/\tau_\textrm{r}^{-1}$), $\kappa_{\text{tot}}$ is the cavity decay rate, $\gamma_{bx}$ is the EEA rate, $\tau_{\text{r}}$ is the total radiative lifetime, $\tau_{nr}$ is the non-radiative lifetime. From now on, $\langle n_\text{c} \rangle$ is noted $\bar{n}_\text{c}$ for convenience. The radiative recombination rate $\tau_{\text{r}}^{-1}\langle N\rangle$ is the sum of emission into two channels: free space and cavity modes. Thus, $\tau_{\text{r}}^{-1} = \tau_{\text{fs}}^{-1} + \tau_{\text{c}}^{-1}$.

Solving for the steady state (and neglecting EEA, i.e. $\gamma_{bx}=0$), the pump rate can be derived:
\begin{equation}
G = \dfrac{\bar{n}_\text{c} \kappa_\text{tot}}{\beta} \left( \dfrac{1 + \beta \bar{n}_\text{c}}{1 + \bar{n}_\text{c}} \right)+R_{\text{defects}}
\end{equation}
where $R_{\text{defects}} = \dfrac{\bar{n}_\text{c} \kappa_\text{tot}}{\beta} \left( \dfrac{1}{1 + \bar{n}_\text{c}} \right) \left( \dfrac{\tau_{\text{r}}}{\tau_{\text{nr}}} \right)$ is the non-radiative recombination rate.

\textit{For an ideal laser} ($\beta=1$), the pump rate needed to satisfy the quantum threshold condition $\bar{n}_c=1$ is:
\begin{equation}
G_{\text{th,ideal}} = \kappa_\text{tot}+R_{\text{defects}} = \kappa_\text{tot}(1+ \dfrac{\tau_{\text{r}}}{\tau_{\text{nr}}}) > \kappa_\text{tot}
\end{equation}
where $R_{\text{defects}}$ can be neglected for a defect-less microlaser, provided that $\dfrac{\tau_{\text{r}}}{\tau_{\text{nr}}}\ll 1$.

It means that the lowest possible threshold to achieve $n_c=1$ in the lasing mode (for an optically pumped ideal single mode laser) $G_{\text{th,ideal}}$ is given by a pump flux that equals or exceeds the cavity decay rate\cite{PhysRevA.50.1675}. For an optically pumped microlaser (assuming $\beta=1$ and neglecting non-radiative losses), the pump rate of the cavity mode given by $G_{\text{th}}=P_{\textrm{th}}/\hbar\omega_{\textrm{p}}$ (where $P_{\textrm{th}}$ is the threshold pump power, $\hbar\omega_{\textrm{p}}$ the pump photon energy) should equal (no defects) or exceed the cavity decay rate (in units $\textrm{s}^{-1}$) given by $\kappa_\text{tot}=\omega_{\textrm{lasing}}/ Q$ (where $\omega_{\textrm{lasing}}$ is the lasing angular frequency and $Q$ the loaded quality factor at the lasing frequency).

\textit{For the less than ideal laser}, accounting for the non-radiative recombination as well as $\beta<1$ and multimodes, we get:
\begin{equation}
G_{\textrm{th}}>M \cdot\kappa_\text{tot}/ QY_\textrm{c}
\label{Gth}
\end{equation}
In particular, below threshold (when stimulated emission does not occur), one has $\bar{n}_\text{c} \approx G \cdot  QY_\text{c}/(M \cdot \kappa_\text{tot})$.

Moreover, the spacing between cavity modes is given by $\Delta f_\textrm{FSR} =\frac{c}{l_\textrm{c}}$, where $l_\textrm{c}$ is the optical path length of the cavity, i.e. for a microdisk $l_\textrm{c} = 2\pi n_{\textrm{g}} R$, where $n_{\textrm{g}}$ is the effective mode index and $R$ is the microdisk radius. The width of emission of the 1L-WSe$_2$ is $\Delta \omega_\textrm{em}=\dfrac{\omega}{Q_\text{em}}$. The total number of modes in the cavity is then:
\begin{equation}
M \approx \dfrac{\Delta \omega_\textrm{em}}{2\pi \Delta f_{FSR}} = \dfrac{l_\textrm{c}}{\lambda Q_\textrm{em}}.
\label{modes}
\end{equation}
The lasing threshold depends \textit{only} on $\kappa_\textrm{tot}$, QY$_c$ and the number of cavity modes $M$, Eq.\ref{Gth}. The cavity finesse is given by
\begin{equation}
\mathcal{F} = \dfrac{FSR}{\Delta\lambda} = \dfrac{\lambda Q}{l_\textrm{c}} = \dfrac{\lambda \omega}{l_\textrm{c}\kappa_\textrm{tot}}
\label{finesse}
\end{equation}
$\kappa_\text{tot}$ is inversely proportional to the cavity length.
From Eqs.\ref{Gth},\ref{modes},\ref{finesse}, we get:
\begin{equation}
G_{th}\sim \dfrac{\omega}{QY_\textrm{c} \cdot Q_\textrm{em} \cdot \mathcal{F}}
\end{equation}
determined only by $QY_\textrm{c}$, $Q_\textrm{em}$, and $\mathcal{F}$. Note that to increase $\mathcal{F}$, one needs to increase its FSR and decrease its mode volume. $\beta$ and Purcell enhancement are extraneous to the task of reducing the threshold an their discussion usually leads to unnecessary confusions. While increasing $\beta$ is often presented as a way to reduce the laser threshold, this is almost always achieved by reducing the active volume of the laser, which keeps the all-important pumping density per unit of volume constant. Similarly, Purcell enhancement does not help, as both stimulated and spontaneous emission rate increase by the same amount\cite{Khurgin2014}.

The determination of the threshold in a microlaser is complicated\cite{Samuel2009,MilonniBook} by the large spontaneous emission coupling factor inherent in the devices with a small volume. The transition from spontaneous to stimulated emission dominated regime occurs in \emph{any laser}, irrespective of implementation, when $\bar{n}_\textrm{c}>1$. This condition has been referred to as the \emph{quantum threshold condition}\cite{PhysRevA.50.1675} in the context of microlasers where it was rigorously shown that when $\bar{n}_c>1$ second order autocorrelation function $g^{(2)}$ approaches unity\cite{Chow2014}, and the laser radiation can be described as a coherent state. Experimentally, the threshold is manifested as narrowing of the emission linewidth by a factor of two as coherent stimulated emission surpasses the incoherent spontaneous emission. Measurement of the linewidth provides the most reliable way to confirm the threshold, but it has often been overlooked when claims of lasing threshold were made based solely on the change of the slope of the input/output curve\cite{Samuel2009,10.1038/nphoton.2017.28}.
\subsection{Cavity quantum yield and $\beta$ factor}
\label{sec:betaQY}
In the simplest model of a multimode cavity, $\textrm{QY}_\textrm{c}$ is defined as:
\begin{equation}
\textrm{QY}_\textrm{c} = \frac{\sum\tau_{c}^{-1}\langle N\rangle}{\sum\tau_{c}^{-1}\langle N\rangle + \tau_{fs}^{-1}\langle N\rangle + \tau_{nr}^{-1}\langle N\rangle}
\label{eq:QYc_def}
\end{equation}
If we call the total radiative contribution $\tau_{r}^{-1}\langle N\rangle = \sum\tau_{c}^{-1}\langle N\rangle + \tau_{fs}^{-1}\langle N\rangle$, dividing the numerator and denominator of Eq.\ref{eq:QYc_def} by this radiative contribution, we get:
\begin{equation}
\textrm{QY}_\textrm{c} = \frac{\sum\beta}{1  + \tau_{r}/\tau_{nr}} = \frac{M\cdot\beta_\text{eff}}{1  + \tau_{r}/\tau_{nr}}
\label{eq:betaQY}
\end{equation}
where $M$ is the number of modes within the envelope of our cavity PL spectrum. Eq.~\ref{eq:betaQY} expresses the relation between the cavity QY and the effective $\beta_\text{eff}$ factor.
\begin{figure}
\includegraphics[width=0.8\columnwidth]{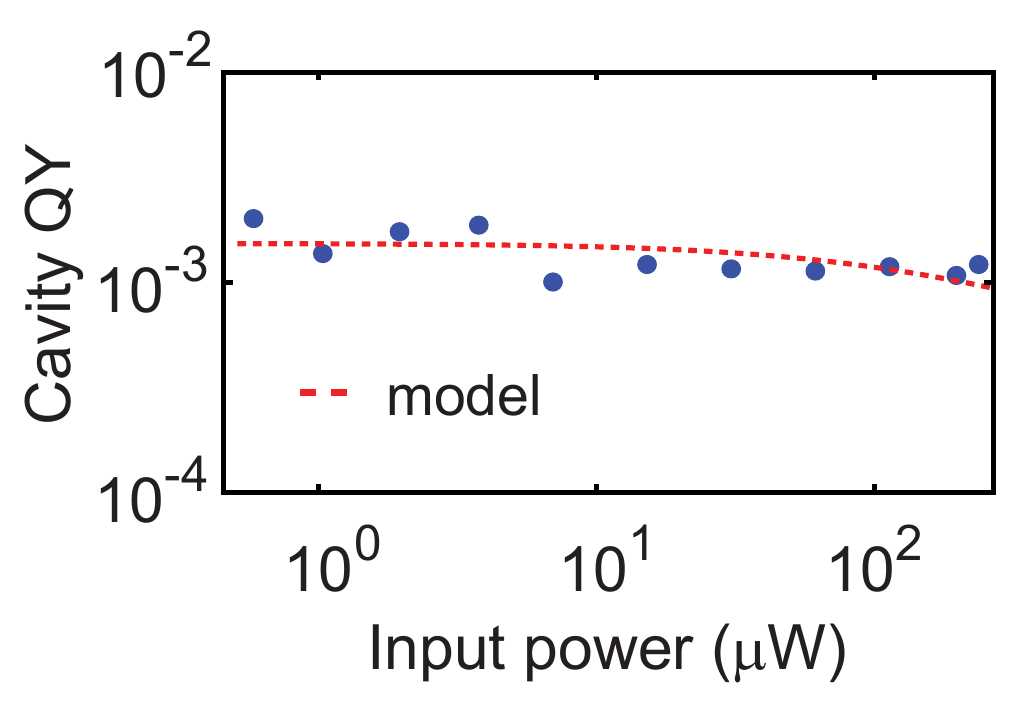}
\caption{QY$_c$ has a nonlinear dependence on excitation input power which is fitted to a model incorporating exciton-exciton annihilation (EEA).}
\label{fig:EEAmodel}
\end{figure}
\subsection{Quantum yield model}
We start with the rate equation for excitons:
\begin{equation}
\frac{d\langle N\rangle}{dt}=\Gamma_p - \Gamma_r\langle N\rangle - \gamma_{bx}\langle N\rangle^2
\label{eq:rateJK2}
\end{equation}
where $\langle N\rangle$ is the number of excitons, $\gamma_{bx}$ is the EEA rate. Note that $\gamma_{bx}$ is not the conventional EEA rate (which has units $\textrm{cm}^2/\textrm{s}$ ), instead it has the unit of $1/\textrm{s}$. Lastly, $\Gamma_r = \tau_{\textrm{fs}}^{-1} + \tau_{\textrm{c,tot}}^{-1} + \tau_{\textrm{nr}}^{-1} =(1+F+F_{\textrm{nr}})\tau_{\textrm{fs}}^{-1}=F_{\textrm{tot}}\tau_{\textrm{fs}}^{-1}$.

Steady state solution of this equation yields:
\begin{equation}
\langle N\rangle = \frac{1}{2\gamma_{bx}}\left(\sqrt{\Gamma_r^2+4\gamma_{bx}\Gamma_p}  - \Gamma_r\right)
\end{equation}
If we call $\Gamma_r^2 / 4\gamma_{bx}$ as $\Gamma_0$ (with $\Gamma_0 = F_{\textrm{tot}}^2\tau_{\textrm{fs}}^{-2}/4\gamma_{\textrm{bx}}$), then the equation is simplified as:
\begin{equation}
\langle N\rangle = \frac{2\Gamma_0}{F_{\textrm{tot}}\tau_{\textrm{fs}}^{-1}}\left(\sqrt{1+\frac{\Gamma_p}{  \Gamma_0}}  - 1\right)
\end{equation}
We then get:
\begin{equation}
QY_c = \frac{\tau_c^{-1}\langle N\rangle}{G} =\frac{F}{F_{\textrm{tot}}} \eta_{\textrm{EEA}}\left(\frac{\Gamma_p}{\Gamma_0}\right)
\end{equation}
where $\eta_{\textrm{EEA}}(x) = 2\frac{\sqrt{1+x} -1 }{x}$.
Expressed in terms of powers, we have $P_{0} = \hbar\omega_p\Gamma_{0}$ and $P = \hbar\omega_p\Gamma_p$. Thus we can fit the following QY expression to two parameters: $\eta_{\textrm{c}}(0)=F/F_{\textrm{tot}} = \tau_{\textrm{c,tot}}^{-1} / (\tau_{\textrm{fs}}^{-1} + \tau_{\textrm{c,tot}}^{-1} + \tau_{\textrm{nr}}^{-1})$, the low power QY$_c$, and $P_{0}$, the power level associated with EEA:
\begin{equation}
QY_c(P) = \eta_{\textrm{c}}(0)\left( \frac{2 (\sqrt{1+P/P_{0} }-1)}{P/ P_{0}  }\right)
\label{eq:QY_model2}
\end{equation}
Fig.\ref{fig:EEAmodel} shows that this model can explain the QY$_c$ excitation power dependence. The fit gives $\eta_{\textrm{c}}(0)=1.5\times 10^{-3}$ and $P_0 = 65~\mu\textrm{W}$. At high power ($>65~\mu W$ in our case), the model predicts a nonlinear dependence of QY$_c$ on pump power.


\begin{thebibliography}{0}%
\makeatletter
\providecommand \@ifxundefined [1]{%
 \@ifx{#1\undefined}
}%
\providecommand \@ifnum [1]{%
 \ifnum #1\expandafter \@firstoftwo
 \else \expandafter \@secondoftwo
 \fi
}%
\providecommand \@ifx [1]{%
 \ifx #1\expandafter \@firstoftwo
 \else \expandafter \@secondoftwo
 \fi
}%
\providecommand \natexlab [1]{#1}%
\providecommand \enquote  [1]{``#1''}%
\providecommand \bibnamefont  [1]{#1}%
\providecommand \bibfnamefont [1]{#1}%
\providecommand \citenamefont [1]{#1}%
\providecommand \href@noop [0]{\@secondoftwo}%
\providecommand \href [0]{\begingroup \@sanitize@url \@href}%
\providecommand \@href[1]{\@@startlink{#1}\@@href}%
\providecommand \@@href[1]{\endgroup#1\@@endlink}%
\providecommand \@sanitize@url [0]{\catcode `\\12\catcode `\$12\catcode
  `\&12\catcode `\#12\catcode `\^12\catcode `\_12\catcode `\%12\relax}%
\providecommand \@@startlink[1]{}%
\providecommand \@@endlink[0]{}%
\providecommand \url  [0]{\begingroup\@sanitize@url \@url }%
\providecommand \@url [1]{\endgroup\@href {#1}{\urlprefix }}%
\providecommand \urlprefix  [0]{URL }%
\providecommand \Eprint [0]{\href }%
\providecommand \doibase [0]{http://dx.doi.org/}%
\providecommand \selectlanguage [0]{\@gobble}%
\providecommand \bibinfo  [0]{\@secondoftwo}%
\providecommand \bibfield  [0]{\@secondoftwo}%
\providecommand \translation [1]{[#1]}%
\providecommand \BibitemOpen [0]{}%
\providecommand \bibitemStop [0]{}%
\providecommand \bibitemNoStop [0]{.\EOS\space}%
\providecommand \EOS [0]{\spacefactor3000\relax}%
\providecommand \BibitemShut  [1]{\csname bibitem#1\endcsname}%
\let\auto@bib@innerbib\@empty
%</preamble>
\end{thebibliography}%


\begin{thebibliography}{62}%
\bibitem{Nanoscale2015} A. C. Ferrari {\em et al.} Science and technology roadmap for graphene, related two-dimensional crystals, and hybrid systems. \href{http://doi.org/10.1039/c4nr01600a}{{\em Nanoscale} \textbf{7}, 4598-4810 (2015)}

\bibitem{10.1038/nmat4205} F. Withers, O. Del Pozo-Zamudio, A. Mishchenko, A. P. Rooney, A. Gholinia, K. Watanabe, T. Taniguchi, S. J. Haigh, A. K. Geim, A. I. Tartakovskii, K. S. Novoselov. Light-emitting diodes by band-structure engineering in Van der Waals heterostructures. \href{http://dx.doi.org/10.1038/nmat4205}{{\em Nat. Mater.} \textbf{14}, 301 (2015)}

\bibitem{/10.1038/nphoton.2016.15} Z. Sun, A. Martinez, F. Wang. Optical modulators with 2d layered materials. \href{http://dx.doi.org/10.1038/nphoton.2016.15}{{\em Nat. Photon.} \textbf{10}, 227 (2016)}

\bibitem{Koppens2014} F. H. L. Koppens, T. Mueller, P. Avouris, A. C. Ferrari, M. S. Vitiello, M. Polini. Photodetectors based on graphene, other two-dimensional materials and hybrid systems. \href{http://dx.doi.org/10.1038/nnano.2014.215}{{\em Nat. Nano.} \textbf{9}, 780 (2014)}

\bibitem{PhysRevB.92.245123} S. Latini, T. Olsen, K. S. Thygesen. Excitons in van der waals heterostructures: The important role of dielectric screening. \href{https://link.aps.org/doi/10.1103/PhysRevB.92.245123}{{\em Phys. Rev. B} \textbf{92}, 245123 (2015)}

\bibitem{PhysRevB.86.115409} A. Ramasubramaniam. Large excitonic effects in monolayers of molybdenum and tungsten dichalcogenides. \href{https://link.aps.org/doi/10.1103/PhysRevB.86.115409}{{\em Phys. Rev. B} \textbf{86} 115409 (2012)}

\bibitem{10.1038/natrevmats.2016.55} J. R. Schaibley, H. Yu, G. Clark, P. Rivera, J. S. Ross, K. L. Seyler, W. Yao, X. Xu. Valleytronics in 2d materials. \href{http://dx.doi.org/10.1038/natrevmats.2016.55}{{\em Nat. Rev. Mat.} \textbf{1}, 16055 (2016)}

\bibitem{PhysRevLett.105.136805} K. F. Mak, C. Lee, J. Hone, J. Shan, T. F. Heinz. Atomically thin ${\mathrm{MoS}}_{2}$: A new direct-gap semiconductor. \href{https://link.aps.org/doi/10.1103/PhysRevLett.105.136805}{{\em Phys. Rev. Lett.} \textbf{105} 136805 (2010)}

\bibitem{10.1038/nphoton.2014.271} F. Xia, H. Wang, D. Xiao, M. Dubey, A. Ramasubramaniam. Two-dimensional material nanophotonics. \href{http://dx.doi.org/10.1038/nphoton.2014.271}{{\em Nat. Photon.} \textbf{8}, 899 (2014)}

\bibitem{Mak2016} K. F. Mak, J. Shan. Photonics and optoelectronics of 2d semiconductor transition metal dichalcogenides. \href{http://dx.doi.org/10.1038/nphoton.2015.282}{{\em Nat. Photon.} \textbf{10}, 216 (2016)}

\bibitem{doi:10.1063/1.4826679} X. Gan, Y. Gao, K. F. Mak, X. Yao, R. J. Shiue, A. van der Zande, M. E. Trusheim, F. Hatami, T. F. Heinz, J. Hone, D. Englund. Controlling the spontaneous emission rate of monolayer MoS$_2$ in a photonic crystal nanocavity. \href{http://dx.doi.org/10.1063/1.4826679}{{\em Appl. Phys. Lett.} \textbf{103} 181119, (2013)}

\bibitem{2053-1583-1-1-011001} S. Wu, S. Buckley, A. M Jones, J. S Ross, N. J Ghimire, J. Yan, D. G Mandrus, W. Yao, F. Hatami, J. Vu\v{c}kovi\'{c}, A. Majumdar, X. Xu. \newblock Control of two-dimensional excitonic light emission via photonic crystal. \href{http://stacks.iop.org/2053-1583/1/i=1/a=011001}{{\em 2D Materials} \textbf{1}, 011001 (2014)}

\bibitem{10.1038/nphoton.2014.304} X. Liu, T. Galfsky, Z. Sun, F. Xia, E. C. Lin, Y. H. Lee, S. K. Cohen, V. M. Menon. Strong light-matter coupling in two-dimensional atomic crystals. \href{http://dx.doi.org/10.1038/nphoton.2014.304}{{\em Nat. Photon.} \textbf{9}, 30 (2015)}

\bibitem{Dufferwiel2015} S. Dufferwiel, S. Schwarz, F. Withers, A. A. P. Trichet, F. Li, M. Sich, O. Del Pozo-Zamudio, C. Clark, A. Nalitov, D. D. Solnyshkov, G. Malpuech, K. S. Novoselov, J. M. Smith, M. S. Skolnick, D. N. Krizhanovskii, A. I. Tartakovskii. Exciton-polaritons in van der waals heterostructures embedded in tunable microcavities. \href{http://dx.doi.org/10.1038/ncomms9579}{{\em Nat. Comm.} \textbf{6}, 8579 (2015)}

\bibitem{10.1038/nphoton.2015.197} Y. Ye, Z. J. Wong, X. Lu, X. Ni, H. Zhu, X. Chen, Y. Wang, X. Zhang. Monolayer excitonic laser. \href{http://dx.doi.org/10.1038/nphoton.2015.197}{{\em Nat. Photon.} \textbf{9}, 733 (2015)}

\bibitem{Amani1065} M. Amani, D.H. Lien, D. Kiriya, J. Xiao, A. Azcatl, J. Noh, S. R. Madhvapathy, R. Addou, Santosh KC, M. Dubey, K. Cho, R. M. Wallace, S.C. Lee, J.H. He, J. W. Ager, X. Zhang, E. Yablonovitch, A. Javey. Near-unity photoluminescence quantum yield in MoS$_2$. \href{http://science.sciencemag.org/content/350/6264/1065}{{\em Science} \textbf{350}, 1065 (2015)}

\bibitem{C5NR00383K} L. Yuan, L. Huang. Exciton dynamics and annihilation in WS$_2$ 2d semiconductors. \href{http://dx.doi.org/10.1039/C5NR00383K}{{\em Nanoscale} \textbf{7}, 7402 (2015)}

\bibitem{doi:10.1021/nl503636c} H. Wang, C. Zhang, F. Rana. Ultrafast dynamics of defect-assisted electron-hole recombination in monolayer MoS$_2$ \href{http://dx.doi.org/10.1021/nl503636c}{{\em Nano Lett.} \textbf{15} 339 (2015)}

\bibitem{10.1038/nprot.2013.087} C. Warth, M. Grabolle, J. Pauli, M. Spieles, U. Resch-Genger. Relative and absolute determination of fluorescence quantum yields of transparent samples. \href{http://dx.doi.org/10.1038/nprot.2013.087}{{\em Nat. Prot.} \textbf{8}, 1535 (2013)}

\bibitem{chizhik2013nanocavity} A. I. Chizhik, I. Gregor, B. Ernst, J. Enderlein. Nanocavity-based determination of absolute values of photoluminescence quantum yields. \href{http:https://doi.org/10.1002/cphc.201200931}{{\em ChemPhysChem} \textbf{14} 505 (2013)}

\bibitem{10.1038/nature14290} S. Wu, S. Buckley, J. R. Schaibley, L. Feng, J. Yan, D. G. Mandrus, F. Hatami, W. Yao, J. Vu\v{c}kovi\'{c}, A. Majumdar, X. Xu. Monolayer semiconductor nanocavity lasers with ultralow thresholds. \href{http://dx.doi.org/10.1038/nature14290}{{\em Nature} \textbf{520}, 69 (2015)}

\bibitem{Li2017} Y. Li, J. Zhang, D. Huang, H. Sun, F. Fan, J. Feng, Z. Wang, Ning C. Z. Room-temperature continuous-wave lasing from monolayer molybdenum ditelluride integrated with a silicon nanobeam cavity. \href{http://dx.doi.org/10.1038/nnano.2017.128}{{\em Nat. Nano.}, advance online publication, (2017)}

\bibitem{Samuel2009} I. D. W. Samuel, E. B. Namdas, G. A. Turnbull. How to recognize lasing. \href{http://dx.doi.org/10.1038/nphoton.2009.173}{{\em Nat. Photon.} \textbf{3} 546 (2009)}

\bibitem{PhysRev.69.674} E. M. Purcell, H. C. Torrey, R. V. Pound. Resonance Absorption by Nuclear Magnetic Moments in a Solid. \href{https://link.aps.org/doi/10.1103/PhysRev.69.674}{{\em Phys. Rev.} \textbf{69}, (1946)}

\bibitem{Haroche1989} S. Haroche, D. Kleppner. Cavity quantum electrodynamics. \href{http://dx.doi.org/10.1063/1.881201}{{\em Physics Today} \textbf{42}, 24 (1989)}

\bibitem{PhysRevB.81.245419} A. Auff\`{e}ves, D. Gerace, J.-M. G\'{e}rard, M. Franca Santos,  L. C. Andreani, J.-P. Poizat. Controlling the dynamics of a coupled atom-cavity system by pure dephasing. \href{https://doi.org/10.1103/PhysRevB.81.245419}{{\em Phys. Rev. B} \textbf{81}, 245419 (2010)}

\bibitem{MilonniBook} P. W. Milonni, J. H. Eberly. {\em Lasers}. Wiley, New York, (1988)

\bibitem{PhysRevA.50.1675} G. Bjork, A. Karlsson, Y. Yamamoto. Definition of a laser threshold, \href{https://link.aps.org/doi/10.1103/PhysRevA.50.1675}{{\em Phys. Rev. A} \textbf{50}, 1675 (1994)}

\bibitem{Chow2014} W. W. Chow, F. Jahnke, C. Gies. Emission properties of nanolasers during the transition to lasing. \href{http://dx.doi.org/10.1038/lsa.2014.82}{{\em Light Sci. Appl.} \textbf{3}, e201 (2014)}

\bibitem{Painter1819} O. Painter, R. K. Lee, A. Scherer, A. Yariv, J. D. O Brien, P. D. Dapkus, I. Kim. Two-dimensional photonic band-gap defect mode laser. \href{http://science.sciencemag.org/content/284/5421/1819}{{\em Science} \textbf{284}, 1819 (1999)}

\bibitem{PhysRevA.74.051802} T. J. Kippenberg, J. Kalkman, A. Polman, K. J. Vahala. Demonstration of an erbium-doped microdisk laser on a silicon chip. \href{http://link.aps.org/doi/10.1103/PhysRevA.74.051802}{{\em Phys. Rev. A} \textbf{74}, 051802, (2006)}

\bibitem{10.1038/nphoton.2017.28} Scrutinizing lasers. \href{http://dx.doi.org/10.1038/nphoton.2017.28}{{\em Nat. Photon.} \textbf{11}, 139 (2017)}

\bibitem{2017arXiv170405341W} G. Wang, C. Robert, M. M. Glazov , F. Cadiz, E. Courtade, T. Amand, D. Lagarde, T. Taniguchi, K. Watanabe, B. Urbaszek, X. Marie.  In-plane Propagation of Light in Transition Metal Dichalcogenide Monolayers: Optical Selection Rules. \href{https://doi.org/10.1103/PhysRevLett.119.047401}{{\em Phys. Rev. Lett.} \textbf{119}, 047401 (2017)}

\bibitem{2017arXiv170105938Z} Y. Zhou, G. Scuri, D. S. Wild , A. A. High, A. Dibos, L. A. Jauregui, C. Shu, K. de Greve, K. Pistunova, A. Joe, T. Taniguchi, K. Watanabe, P. Kim, M. D. Lukin, H. Park. Probing dark excitons in atomically thin semiconductors via near-field coupling to surface plasmon polaritons. \href{https://doi.org/10.1038/nnano.2017.106}{{\em Nat. Nano.} \textbf{12}, 856–860 (2017)}

\bibitem{Heterogeneousintegration} P. Rojo Romeo, J. Van Campenhout, P. Regreny, A. Kazmierczak, C. Seassal, X. Letartre, G. Hollinger, D. Van Thourhout, R. Baets, J.M. Fedeli, and L. Di Cioccio. Heterogeneous integration of electrically driven microdisk based laser sources for optical interconnects and photonic ICs. \href{https://doi.org/10.1364/OE.14.003864}{{\em Opt. Express} \textbf{14}, 3864-3871 (2006)}

\bibitem{PhysRevApplied.5.054019} R. Schilling, H. Sch\"utz, A. H. Ghadimi, V. Sudhir, D. J. Wilson, T. J. Kippenberg. Near-field integration of a sin nanobeam and a ${\mathrm{SiO}}_{2}$ microcavity for heisenberg-limited displacement sensing. \href{https://link.aps.org/doi/10.1103/PhysRevApplied.5.054019}{{\em Phys. Rev. Appl.} \textbf{5}, 054019 (2016)}

\bibitem{GHULINYAN2015679} M. Ghulinyan, M. Bernard, R. Bartali, G. Pucker. Formation of mach angle profiles during wet etching of silica and silicon nitride materials. \href{http://dx.doi.org/10.1016/j.apsusc.2015.10.114}{{\em Appl. Surf. Sci.} \textbf{359}, 679 (2015)}

\bibitem{10.1038/nature01939} K. J. Vahala. Optical microcavities. \href{http://dx.doi.org/10.1038/nature01939}{{\em Nature} \textbf{424}, 839 (2003)}

\bibitem{10.1038/ncomms12978} C. Palacios-Berraquero, M. Barbone, D. M. Kara, X. Chen, I. Goykhman, D. Yoon, A. K. Ott, J. Beitner, K. Watanabe, T. Taniguchi, A. C. Ferrari, M. Atat\"{u}re. Atomically thin quantum light-emitting diodes. \href{http://dx.doi.org/10.1038/ncomms12978}{{\em Nat. Comm.} \textbf{7}, 12978 (2016)}

\bibitem{10.1021/nl071168m} C. Casiraghi, A. Hartschuh, E. Lidorikis, H. Qian, H. Harutyunyan, T. Gokus, K. S. Novoselov, A. C. Ferrari. Rayleigh imaging of graphene and graphene layers. \href{http://doi.org/10.1021/nl071168m}{{\em Nano Lett.} \textbf{7}, 2711–2717 (2007)}

\bibitem{2053-1583-1-1-011002} A. Castellanos-Gomez, M. Buscema, R. Molenaar, V. Singh, L. Janssen, H. S. J. Van der Zant, G. A. Steele. Deterministic transfer of two-dimensional materials by all-dry viscoelastic stamping. \href{http://stacks.iop.org/2053-1583/1/i=1/a=011002}{{\em 2D Materials} \textbf{1} 011002, (2014)}

\bibitem{doi:10.1038/srep04215} H. Terrones, E. Del Corro, S. Feng, J. M. Poumirol, D. Rhodes, D. Smirnov, N. R. Pradhan, Z. Lin, M. A. T. Nguyen, A. L. El\'{i}as, T. E. Mallouk, L. Balicas, M. A. Pimenta, M. Terrones. New first order Raman-active modes in few layered transition metal dichalcogenides. \href{https://doi.org/10.1038/srep04215}{{\em Sci. Rep.} \textbf{4}, 4215 (2014)}

\bibitem{10.1039/C3NR03052K} W. Zhao, Z. Ghorannevis, K. K. Amara, J. R. Pang,M. Toh, X. Zhang,C. Kloc, P. H. Tan, G. Eda. Lattice dynamics in mono- and few-layer sheets of WS$_2$ and WSe$_2$. \href{https://doi.org/10.1039/C3NR03052K}{{\em Nanoscale} \textbf{5}, 9677–83 (2013)}

\bibitem{10.1103/PhysRevB.90.205422} Y. Li, A. Chernikov, X. Zhang, A. Rigosi, H. M. Hill, A. M. van der Zande, D. A. Chenet, E.-M. Shih, J. Hone, T. F. Heinz. Measurement of the optical dielectric function of monolayer transition-metal dichalcogenides: ${\mathrm{MoS}}_{2}$, $\mathrm{Mo}\mathrm{S}{\mathrm{e}}_{2}$, ${\mathrm{WS}}_{2}$, and $\mathrm{WS}{\mathrm{e}}_{2}$. \href{https://doi.org/10.1103/PhysRevB.90.205422}{{\em Phys. Rev. B} \textbf{90}, 205422 (2014)}

\bibitem{PhysRevLett.91.043902} S. M. Spillane, T. J. Kippenberg, O. J. Painter, K.J. Vahala. Ideality in a fiber-taper-coupled microresonator system for application to cavity quantum electrodynamics. \href{https://link.aps.org/doi/10.1103/PhysRevLett.91.043902}{{\em Phys. Rev. Lett.} \textbf{91} 043902 (2003)}

\bibitem{PhysRevLett.85.74} M. Cai, O. Painter, K. J. Vahala. Observation of critical coupling in a fiber taper to a silica-microsphere whispering-gallery mode system. \href{https://link.aps.org/doi/10.1103/PhysRevLett.85.74}{{\em Phys. Rev. Lett.} \textbf{85}, 74 (2000)}

\bibitem{Haus1984} H. A. Haus. {\em Waves and fields in optoelectronics}. Prentice Hall, (1984)

\bibitem{Weiss:95} D. S. Weiss, V. Sandoghdar, J. Hare, V. Lef\`{e}vre-Seguin, J. M. Raimond,  S.Haroche. Splitting of high-Q Mie modes induced by light backscattering in silica microspheres. \href{http://ol.osa.org/abstract.cfm?URI=ol-20-18-1835}{{\em Opt. Lett.} \textbf{20} 1835 (1995)}

\bibitem{doi:10.1021/acs.nanolett.5b01665} O. Salehzadeh, M. Djavid, N. H. Tran, I. Shih, Z. Mi. Optically pumped two-dimensional MoS$_2$ lasers operating at room-temperature. \href{http://dx.doi.org/10.1021/acs.nanolett.5b01665}{{\em Nano Lett.} \textbf{15}, 5302 (2015)}

\bibitem{doi:10.1021/nl503312x} S. Schwarz, S. Dufferwiel, P. M. Walker, F. Withers, A. A. P. Trichet, M. Sich, F. Li, E. A. Chekhovich, D. N. Borisenko, N. N. Kolesnikov, K. S. Novoselov, M. S. Skolnick, J. M. Smith, D. N. Krizhanovskii, A. I. Tartakovskii. Two-dimensional metal chalcogenide films in tunable optical microcavities. \href{http://dx.doi.org/10.1021/nl503312x}{{\em Nano Lett.} \textbf{14}, 7003 (2014)}

\bibitem{PhysRevLett.113.026803} K. He, N. Kumar, L. Zhao, Z. Wang, K. F. Mak, H. Zhao, J. Shan. Tightly bound excitons in monolayer ${\mathrm{WSe}}_{2}$. \href{https://link.aps.org/doi/10.1103/PhysRevLett.113.026803}{{\em Phys. Rev. Lett.} \textbf{113} 026803, (2014)}

\bibitem{Grillet:07} C. Grillet, C. Monat, C. L. C. Smith, B. J. Eggleton, D. J. Moss, S. Fr\'{e}d\'{e}rick, D. Dalacu, P. J. Poole, J. Lapointe, G. Aers, R. L. Williams. Nanowire coupling to photonic crystal nanocavities for single photon sources. \href{http://www.opticsexpress.org/abstract.cfm?URI=oe-15-3-1267}{{\em Opt. Expr.} \textbf{15}, 1267 (2007)}

\bibitem{PSSB:PSSB201600563} N. B. Mohamed, F. Wang, H. E. Lim, W. Zhang, S. Koirala, S. Mouri, Y. Miyauchi, K. Matsuda. Evaluation of photoluminescence quantum yield of monolayer WSe$_2$ using reference dye of 3-borylbithiophene derivative. \href{http://dx.doi.org/10.1002/pssb.201600563}{{\em Phys. Stat. Sol. (b)} \textbf{254}, 160056 (2017)}

\bibitem{PhysRevB.90.075413} G.~Wang, L.~Bouet, D.~Lagarde, M.~Vidal, A.~Balocchi, T.~Amand, X.~Marie, B.~Urbaszek. Valley dynamics probed through charged and neutral exciton emission in monolayer ${\mathrm{WSe}}_{2}$. \href{https://link.aps.org/doi/10.1103/PhysRevB.90.075413}{{\em Phys. Rev. B} \textbf{90} 075413 (2014)}

\bibitem{doi:10.1063/1.4895471} T. Yan, X. Qiao, X. Liu, P. H. Tan, X. Zhang. Photoluminescence properties and exciton dynamics in monolayer WSe$_2$. \href{http://dx.doi.org/10.1063/1.4895471}{{\em Appl. Phys. Lett.} \textbf{105}, 101901 (2014)}

\bibitem{purcell1995spontaneous} E. M. Purcell. Spontaneous emission probabilities at radio frequencies. \href{https://doi.org/10.1007/978-1-4615-1963-8_40}{{\em Confined Electrons and Photons}, 839-839. Springer (1995)}

\bibitem{PhysRevLett.116.247402} A. Jeantet, Y. Chassagneux, C. Raynaud, P. Roussignol, J. S. Lauret, B. Besga, J. Est\`eve, J. Reichel, C. Voisin. Widely tunable single-photon source from a carbon nanotube in the Purcell regime. \href{https://link.aps.org/doi/10.1103/PhysRevLett.116.247402}{{\em Phys. Rev. Lett.} \textbf{116} 247402 (2016)}

\bibitem{refId0} {Gayral, B.} Controlling spontaneous emission dynamics in semiconductor microcavities. \href{https://doi.org/10.1051/anphys:200102001}{{\em Ann. Phys. Fr.} \textbf{26}, 1 (2001)}

\bibitem{Meldrum:10} A. Meldrum, P. Bianucci, F. Marsiglio. Modification of ensemble emission rates and luminescence spectra for inhomogeneously broadened distributions of quantum dots coupled to optical microcavities. \href{http://www.opticsexpress.org/abstract.cfm?URI=oe-18-10-10230}{{\em Opt. Expr.} \textbf{18}, 10230 (2010)}

\bibitem{PhysRevB.90.155449} S. Mouri, Y. Miyauchi, M. Toh, W. Zhao, G. Eda, K. Matsuda. Nonlinear photoluminescence in atomically thin layered ${\mathrm{WSe}}_{2}$ arising from diffusion-assisted exciton-exciton annihilation. \href{https://link.aps.org/doi/10.1103/PhysRevB.90.155449}{{\em Phys. Rev. B} \textbf{90}, 155449 (2014)}

\bibitem{doi:10.1063/1.4978868} M. J. Yang, C. C. Lin, Y.S. Wu, L. Wang, N. Na. Optical properties of organic-silicon photonic crystal nanoslot cavity light source. \href{http://dx.doi.org/10.1063/1.4978868}{{\em AIP Advances} \textbf{7}, 035309 (2017)}

\bibitem{Khurgin2014} J. B. Khurgin, G. Sun. Comparative analysis of spasers, vertical-cavity surface-emitting lasers and surface-plasmon-emitting diodes. \href{http://dx.doi.org/10.1038/nphoton.2014.94}{{\em Nat. Photon.} \textbf{8}, 468 (2014)}

\bibitem{10.1088/2053-1583/aa6aa1} O. Ajayi, J. Ardelean, G. Shepard, J. Wang, A. Antony, T. Taniguchi, K. Watanabe, T. Heinz, S. Strauf, X.Y. Zhu, J. C. Hone. Approaching the intrinsic photoluminescence linewidth in transition metal dichalcogenide monolayers. \href{http://iopscience.iop.org/10.1088/2053-1583/aa6aa1}{{\em 2D Materials} \textbf{4}, 3 (2017)}

\bibitem{10.1103/PhysRevB.93.201111} Y. Yu, Y. Yu, C. Xu, A. Barrette, K. Gundogdu, L. Cao. Fundamental limits of exciton-exciton annihilation for light emission in transition metal dichalcogenide monolayers. \href{https://doi.org/10.1103/PhysRevB.93.201111}{{\em Phys. Rev. B} \textbf{93}, 201111 (2016)}

\end{thebibliography}
\end{document}